\documentclass[aoas,nameyear,dvips]{arximspdf}
\usepackage{dcolumn}
\usepackage{graphicx}

\doi{10.1214/09-AOAS252}
\volume{3}
\issue{4}
\pubyear{2009}
\firstpage{1655}
\lastpage{1674}

\makeatletter
\newcolumntype{d}[1]{D{.}{.}{#1}}
\makeatother

\begin{document}
\begin{frontmatter}

\title{Nonparametric Bayesian multiple testing for longitudinal
performance stratification\thanksref{T1}}
\runtitle{Nonparametric multiple testing}
\thankstext{T1}{Supported by a graduate research fellowship from the
U.S.~National Science Foundation.}

\begin{aug}
\author{\fnms{James G.} \snm{Scott}\ead[label=e1]{james.scott@mccombs.utexas.edu}\corref{}}
\address{University of Texas at Austin\\McCombs School of Business\\
1 University Station, B6500\\Austin, Texas 78712\\ USA \\ \printead{e1}}
\affiliation{University of Texas at Austin}
\runauthor{J. G. Scott}
\end{aug}

\received{\smonth{4} \syear{2008}}
\revised{\smonth{4} \syear{2009}}

%
\begin{abstract}
This paper describes a framework for flexible multiple hypothesis
testing of autoregressive time series. The modeling approach is
Bayesian, though a blend of frequentist and Bayesian reasoning is used
to evaluate procedures. Nonparametric characterizations of both the
null and alternative hypotheses will be shown to be the key
robustification step necessary to ensure reasonable Type-I error
performance. The methodology is applied to part of a large database
containing up to 50 years of corporate performance statistics on 24,157
publicly traded American companies, where the primary goal of the
analysis is to flag companies whose historical performance is
significantly different from that expected due to chance.
\end{abstract}

%
\begin{keyword}
\kwd{Multiple testing}
\kwd{Bayesian model selection}
\kwd{nonparametric Bayes}
\kwd{financial time series}.
\end{keyword}

\end{frontmatter}
%

\section{Introduction}
\label{intro}

\subsection{Multiple testing of time series}\label{sec1.1}

Suppose a single time series of length $T$ is observed for each of $N$
different units. Two possible models for each time series are
entertained: a simple autoregressive null model $M_0$ and a more
complex alternative model $M_A$. The goal is to determine which units
come from the alternative model.

This is a common problem in the analysis of multiple time series, and
although many details will be context-dependent, certain common themes
emerge. Of key interest is how, in repeatedly applying a procedure used
for testing a single time series, the rate of Type-I errors can be
controlled. Model-based approaches are an attractive option, but model
errors can become overwhelming in the face of massive multiplicity. One
of this paper's main results is that great care must be taken in
characterizing $M_0$ and $M_A$ in order to keep false positives at bay,
with the suggested robustification step involving the use of
nonparametric Bayesian methods.

In typical hypothesis-testing scenarios involving standard parametric
models $M_0$ and $M_A$, the Bayes factor $\operatorname{BF}(M_A \dvtx  M_0)$
contains an ``Ockham's razor'' term [\citet{jefferysberger92}] that
penalizes the more complex model. In most cases this is the result of
needing to integrate the likelihood across a higher-dimensional prior
under the more complex model, which will therefore have a more diffuse
predictive distribution. This penalty for model complexity is quite
different from any sort of penalty term imposed for conducting multiple
hypothesis tests [\citet{scottberger2007}].

But the multiple-testing framework employed here uses Dirichlet-process
mixtures to represent the unknown null and alternative models, which
are properly thought of as being infinite-dimensional. Another of this
paper's main objectives is to clarify the nature and operating
characteristics of this penalty for model complexity in nonparametric
settings---specifically, how it interfaces with the recommended
multiple-testing methodology.

\subsection{Motivating example}\label{sec1.2}

A running example from management theory will be used to motivate and
study the proposed Bayesian model. The results and discussion are
problem-specific, but areas in which the methodology and lessons can be
generalized will be pointed out.

Our data set covers up to 50 years of annual performance
(operationalized by a common accounting measure) for over 24,000
publicly traded American companies. The goal of the analysis is to flag
firms whose performance is highly unlikely to have occurred by random
chance, since these firms may have good (or bad) management practices
that are discernible through follow-up case studies.

Longitudinal performance stratification is a classic topic in
management theory. Indeed, one of the primary aims of
strategic-management research, and the conceit of many best-selling
books, is to explain why some firms fail and others succeed.

Much academic work in this direction focuses on decomposing observed
variation into market-level, industry-level and firm-level
components\break
[\citet{bowmanhelfat2001}, \citet{hawawinietal2003}]. Of the work that
attempts to identify specific nonrandom performers, much of it relies
upon model-free clustering algorithms [e.g., \citet{harrigan1985}], which contain no guarantee that the clusters found
will be significantly different from one another. Other approaches
employ simple classical tests [\citet{ruefliwiggins2000}], often
based upon ordinal time series. These have the advantage of being
model-free, but are typically not based upon sufficient statistics, and
suffer from the fact that available multiplicity-correction approaches
(e.g.,~Bonferroni correction) tend toward the overly aggressive.

Due to the number of firms for which public financial data is
available, this problem makes an excellent testbed for the study of
general-purpose multiple-testing methodology in time-series analysis.
There is, however, no theoretical ideal of what an
``average-performing'' company should look like, beyond the notion that
it should revert to the population-level mean even if it has some
randomly good or bad years. The Bayesian approach requires that
suitable notions of randomness and nonrandomness be embodied in a
statistical model. This model must confront an obvious multiplicity
problem: many thousands of companies will be tested, and false
positives will make expensive wild-goose chases out of any follow-up
studies seeking to explain possible sources of competitive advantage.
Robustness and trustworthy Type-I error characteristics are therefore
crucial practical considerations, and so even though this paper's
modeling approach is Bayesian, it also contains much frequentist
reasoning regarding Type-I error rates.

\section{Testing a zero-mean AR(1) model}
\label{basicmodel}

\subsection{The model}\label{sec2.1}

This section outlines a basic framework for multiple testing that, for
the sake of illustration, will be purposely simplistic. Nonetheless, it
will provide a useful jumping-off point for the methodological
developments of subsequent sections, and will show why more flexible
models are typically needed in order to achieve reasonable Type-I error
performance.

Let $y_{it}$ be the observation for unit $i$ at time $t$, and let
$\mathbf{y}_i$ be the vector of observations for unit $i$. In the
management-theory example, $y$ is a standardized performance metric
called Return on Assets (ROA), which measures how efficiently a
company's assets generate earnings. Each company's ROA values were
regressed upon a set of covariates judged to be relevant by three
experts in management theory collaborating on the project. These
include the company's size, debt-to-equity ratio and market share,
along with categorical variables for year and for industry membership.
[See \citet{ruefliwiggins2002}, for a summary of the literature regarding covariate
effects on observed firm performance.] The actual
values used in the following analyses were the residuals from this
regression. Also, since the question at issue is one of relative
performance, not absolute performance, these residuals were
standardized by CDF transform to follow a $\mathrm{N}(0,1)$ distribution.

Since we do not expect random gains or losses in one year to be
completely erased by the following year, a model accounting for serial
autocorrelation seems mandatory. Management-theoretic support for this
assumption in the present context can be found in \citet{denrell2003}
and \citet{denrell2005}; analogous situations in engineering, finance
and biology are very common.

The null hypothesis is then a stationary AR(1) model depending upon
parameter $\bolds{\theta} = (\phi, v)$:
\begin{eqnarray*}
y_{it} &=& \phi y_{i,(t-1)} + \nu_{it}, \\
\nu_{it} &\stackrel{\mathrm{i.i.d.}}{\sim}& \mathrm{N}(0, v)   .
\end{eqnarray*}
This assumption allows for long runs of good or bad performance due
simply to chance: a large shock ($\nu_t$) may take quite awhile to
decay depending upon the value of $\phi$, which is assumed to lie on $(-1,1)$.

Nonnull companies can then be modeled as AR(1) processes that revert
to a nonzero mean. Placing a mixture prior on this unknown mean will
then encode the relevant hypothesis test:
%
\begin{eqnarray}
\mathbf{y}_{i} &\sim&\mathrm{N}(\mathbf{y}_{i} \mid m_i \mathbf{1},
\Sigma
_{\theta}), \label{simple1} \\
\bolds{\theta} &\sim&\mathrm{N}(\phi\mid d, D) \times
\operatorname{IG}(v \mid a,
b), \label{simple2} \\
m_i &\sim& p \cdot\mathrm{N}(0,\sigma^2) + (1-p) \delta_0 \label{simple3}
  ,
\end{eqnarray}
with $\mathbf{1}$ is the vector of all ones, $\Sigma_{\theta}$ is
the familiar AR(1) variance matrix, $\delta_0$ is a point mass at 0,
and $p \in[0,1]$ is the prior probability of arising from the
alternative hypothesis. The exchangeable normal prior on the nonzero
means $m_i$ reflects the prior belief that, among firms that
systematically deviate from zero, most of the deviations will be
relatively small.

The posterior probabilities $p_i = P(m_i \neq0 \mid Y)$ then can be
used to flag nonnull units. In some contexts, these are called the
posterior inclusion probabilities, reflecting inclusion in the
``nonnull'' set.

The model must be completed by specifying priors for $\phi$, $v$ and
$\sigma^2$. In the example at hand, the data have been standardized to
a $\mathrm{N}(0,1)$ scale, so it makes sense to choose $\sigma^2 =
1$. In
more general settings, the prior for $\sigma^2$ must be appropriately
scaled by $\phi$ and $v$ in the absence of strong prior information,
since the marginal variance of the residual autoregressive process is
the only quantity that provides a default scale for the problem.

The conditional likelihoods of each data vector under the two
hypotheses (\mbox{$m_i \neq0 $} versus $m_i = 0$) are available in closed form:
%
\begin{eqnarray}
P(\mathbf{y}_i \mid m_i = 0, \bolds{\theta}) &=& \mathrm
{N}
(\mathbf{y}_i \mid\mathbf{0}, \Sigma_{\theta} ), \label
{condlike1} \\
P(\mathbf{y}_i \mid m_i \neq0, \bolds{\theta}, \sigma^2) &=&
\mathrm{N}\bigl(\mathbf{y}_i \mid\mathbf{0}, \Sigma_{\theta} +
\sigma^2
(\mathbf{1} \mathbf{1}^t) \bigr) \label{condlike2}   ,
\end{eqnarray}
where $(\mathbf{1} \mathbf{1}^t)$ is the matrix of all ones.

The ratio of (\ref{condlike2}) to (\ref{condlike1}) gives the Bayes
factor, conditional upon $\phi$, $v$ and $\sigma^2$, for testing an
individual time series against the null model.

\subsection{Bayesian adjustment for multiplicity}\label{sec2.2}

Multiplicity, from a Bayesian perspective, is handled through careful
treatment of the prior probability $p$.

One possibility is to choose a small value of $p$, with the expectation
that a prior bias in favor of the null will solve the problem. But the
key intuition in using this model for multiple testing is to treat $p$
as just another model parameter to be estimated from the data, giving
it a uniform prior. This induces an effect that is often referred to as
an automatic multiple-testing penalty, with the effect being
``automatic'' in the sense that no arbitrary penalty terms must be specified.

This effect can most easily be seen if one imagines repeatedly testing
a fixed number of signals in the presence of an increasing number of
null units. Asymptotically, the posterior mass of $p$ will concentrate
near $0$, making it increasingly difficult for all units (even the
signals) to overcome the prior belief in their irrelevance. This yields
much the same effect as choosing a small value for $p$ after the fact,
but Bayesian learning about $p$ negates the need to make such an
arbitrary choice.

Two caveats are in order. First, this should not be misinterpreted as
saying that Bayesians need never worry about multiplicities. Automatic
adjustment depends upon allowing the data itself to choose $p$, and
more generally, upon careful treatment of prior model probabilities.
The adjustment itself can be seen primarily in the inclusion
probabilities, and will not necessarily be incorporated into other
posterior quantities of interest. In particular, the joint distribution
of effect sizes, conditional on their being nonzero, will fail to
adjust for multiplicity.

Second, it is still necessary to choose a threshold on posterior
probabilities in order to get a decision rule about ``signal'' versus
``noise.'' An obvious threshold is $50\%$, but ideally this threshold
should be chosen to minimize Bayesian expected loss under properly
specified loss functions. Perhaps, for example, the loss incurred by a
false positive is constant while the loss incurred by a false negative
scales according to some function of the underlying difference from
$0$. Introducing such loss functions will complicate the analysis only
slightly, without changing the fundamental need to account for multiplicity.

It is important to distinguish this fully Bayesian model from the
empirical-Bayes approach to multiplicity correction, whereby $p$ is
estimated by maximum likelihood [see, e.g., \citet
{johnstonesilverman2004}]. These two approaches are intuitively
similar, since the prior inclusion probability is estimated by the data
in order to yield an automatic multiple-testing penalty. Yet the
approaches have quite different operational and theoretical properties
[\citet{scottberger2007}], with the focus here being on the fully
Bayesian approach.

Similar multiple-testing procedures have been extensively studied in
many different contexts. See \citet{scottberger06} for theoretical
development; \citet{domuller2005} for genomics; \citet
{scottcarvalho2007b} and \citet{carvalhoscott2007} for portfolio
selection; and \citet{georgefoster2000} and \citet{cuigeorge2006} for
examples in regression.

The posterior inclusion probabilities $\{p_i\}$ can be computed
straightforwardly using importance sampling to account for posterior
uncertainty about $\phi$, $v$ and $p$, which will be stable as long as
$p$ is not too close to $0$ or $1$. The advantage of importance
sampling here is that a common importance function may be used for all
marginal inclusion probabilities, greatly simplifying the required
calculations. After transforming all variables to have unrestricted
domains, importance sampling was used to compute the results presented
in this section, with repetition and plots of the importance weights
used to confirm stability.

\subsection{Results on ROA data}
\label{basicmodelresults}

Historical ROA time series for 3459 publicly traded American companies
between 1965 and 2004 were used to fit the above model. This
encompasses almost every public company over that period for which at
least 20 years of data were available. Standard independent conjugate
priors for $\phi$ and $v$ were used:
%
\begin{eqnarray}
\phi&\sim&\mathrm{N}_U(0.5,0.25^2) \label{basicprior1}, \\
v &\sim&\operatorname{IG}(2,1) \label{basicprior2}   ,
\end{eqnarray}
where $\mathrm{N}_U$ indicates that the normal prior for $\phi$ is truncated
to lie on $(-1, 1)$. These priors were chosen to reflect the
expectations of the collaborating management theorists regarding the
persistence and scale of random ROA fluctuations from year to year.
Because these parameters appear in both the null and alternative
models, integrals over these priors appear in both the numerator and
denominator of the Bayes factor comparing $m_i \neq0$ versus $m_i =
0$, making them not overly influential in the analysis; see \citet
{bergervarsh1998} for general guidelines on choosing common
hyperparameters in model-selection problems.

\begin{figure}

\includegraphics{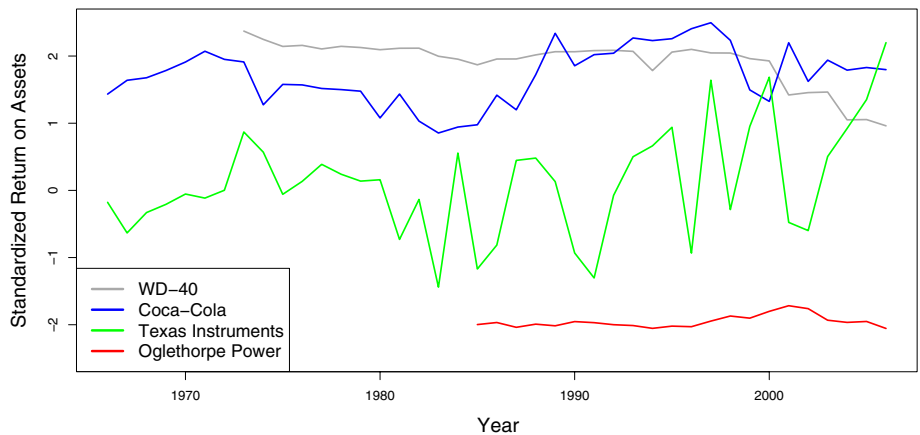}

\caption{Four example companies.}
\label{examplefirms}
\end{figure}

In many cases the results of the fit seemed reasonable. Most firms were
assigned to $M_0$ with high probability, while companies with obvious
patterns of sustained excellence or inferiority were flagged as being
from $M_A$ with very high probability. Figure \ref{examplefirms}
contains instructive examples: two excellent companies (WD-40 and
Coca-Cola), along with one obviously poor company (Oglethorpe Power),
were assigned greater than $95\%$ probability of being nonnull. A
fourth example, Texas Intruments, had several intermittent years of
good performance but no pattern of sustained excellence, and the model
gave it greater than $90\%$ probability of being from the null model.

\begin{figure}

\includegraphics{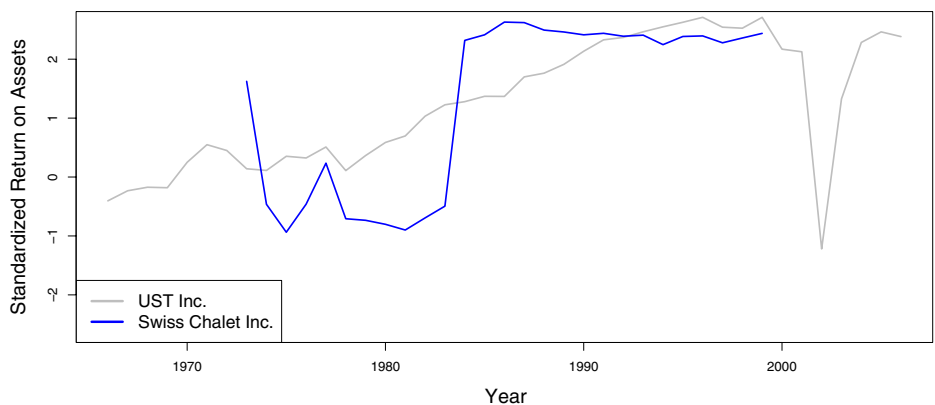}

\caption{Two examples that seem to display trends.}
\label{trendexamples}
\end{figure}

On the other hand, the model displayed two serious shortcomings:
\begin{itemize}
\item Many firms diverged in obvious ways (e.g., via the
appearance of long-term trends) from the expectations of a single AR(1)
model. Two such examples are in Figure \ref{trendexamples}. Discussion
of this important issue is postponed until Section~\ref{DPGP}.

\item More subtly, the model imposed a homogeneous error structure on
data that seemed rather heterogeneous. Some fairly basic exploratory
data analysis indicated that firms displayed differing degrees of
``stickiness'' in their trajectories. This suggested that a single
value of $\phi$ for the entire data set might be unsatisfactory.
Likewise, some firms appeared systematically more volatile than others,
making a single-variance model equally questionable.
\end{itemize}

\subsection{Robustness simulations}
\label{prelimrobustnessstudy}

The possibility of model errors in (\ref{simple1})--(\ref{simple3})
bring the issue of robustness to the forefront. This section describes
the results of a simulation study that shows just how poorly this model
can perform when a particular type of model error is encountered:
deviation from the ``single $\phi$, single $v$'' approach to
describing the AR(1) residuals of all companies in the sample.

Several data sets displaying different levels of heterogeneity were
simulated. The homogeneous (i.e.,~single $\phi$, single $v$) model was
subsequently fit to each simulated data set in order to assess the
robustness of the procedure's Type-I error performance.

Each simulated data set had 3500 times series of length $T=40$, with
each time series drawn from a mixture distribution of AR(1) models.
These distributions ranged from trivial one-component mixtures (for
which the assumed model was true) to complex nine-component mixtures
(for which the assumed model was quite a bad approximation). These
conditions are summarized in Table \ref{robustnesstable}. In the four-
and nine-component models, all components were equiprobable. Since all
simulated companies had $m_i = 0$, ideally there should be no positive flags.

For the purposes of classification, thresholding is reported at the
$p_i \geq0.5$ and the $p_i \geq0.9$ levels, where $p_i$ is the
posterior inclusion probability for company~$i$. The first ($p_i \geq
0.5$) threshold reflects a 0--1 loss function that symmetrically
penalizes false positives and false negatives. The second threshold is
arbitrary, but meant to reflect a more conservative approach to
identifying signals. A full decision-theoretic analysis incorporating
more realistic loss functions would yield a different, data-adaptive threshold.

Table \ref{robustnesstable} supports two conclusions:
\begin{itemize}
\item The proposed model yields very strong control over false
positives when its assumptions are met: 3 false positives and 0 false
positives in the two cases investigated, out of 3500 units tested.
This confirms that the theory outlined in \citet{scottberger06}, which
concerns the much simpler normal-means testing problem, applies here as well.
\item This excellent Type-I error profile is not at all robust to a
violation of the autoregressive model's assumptions. In the most
extreme case, nearly a third of units (1045 out of 3500) tested had
inclusion probabilities $p_i \geq50\%$, when in reality none were from
the alternative model. In other less extreme cases, the false positives
still numbered in the hundreds, which is clearly unsatisfactory.
\end{itemize}

\begin{table}
\caption{Robustness of the multiple-testing
procedure's Type-I error performance to heterogeneity in the
autoregressive profiles of tested units. Here $\hat{p}$ refers to the
posterior mode of the mixing ratio $p$, and~the $p_i$'s are the
posterior probabilities that each $m_i \neq0$}\label{robustnesstable}
\begin{tabular*}{\textwidth}{@{\extracolsep{\fill}}lcd{4.0}d{3.0}@{}}
\hline
\textbf{Number of components: Model} & $\bolds{\hat{p}}$ & \multicolumn{1}{c}{\textbf{\#} $\bolds{p_i \geq0.5}$}
& \multicolumn{1}{c@{}}{\textbf{\#} $\bolds{p_i\geq0.9}$} \\
\hline
1: $(\phi,v) = (0.5, 0.25)$ & 0.01 & 3 & 0 \\
1: $(\phi,v) = (0.9, 0.5)$ & 0.02 & 0 & 0 \\
4: $(\phi, v) \in\{0.5, 0.7\} \times\{0.25,0.5\}$ & 0.02 & 30 & 4 \\
4: $(\phi, v) \in\{0.4, 0.6, 0.8\} \times\{0.05,0.25,0.5\}$ & 0.08 &152 & 66 \\
9: $(\phi, v) \in\{0.2, 0.95\} \times\{0.05,0.5\}$ & 0.37 & 1045 &560 \\
9: $(\phi, v) \in\{0.2, 0.6, 0.95\} \times\{0.05,0.25,0.5\}$ & 0.29& 797 & 493 \\
\hline
\end{tabular*}
\end{table}

These results dramatically illustrate the effect of heterogeneity in
the autoregressive profiles of each tested unit. If such heterogeneity
exists but is ignored, the Type-I error performance of the procedure
may be severely compromised.

\section{A nonparametric null model}
\label{DPAR}

In the previous section, a specific form of model error---different
groups of companies following different AR models---was shown to be a
source of overwhelming Type-I errors. Hence, a natural extension is to
consider a more complicated autoregressive model for the residuals that
accounts for the possibility of stratification.

The Dirichlet process [\citet{ferguson1973}] offers a straightforward
nonparametric technique for accommodating uncertainty about this random
distribution. Let $\mathbf{z}_i$ represent the response vector for
unit $i$, for now ignoring any contribution due to a nonzero mean.
Recall that for parameter $\bolds{\theta} = (\phi, v)$, $\Sigma
_{\theta}$ denotes the AR(1) variance matrix. The DP mixture model can
then be written as a hierarchical model:\looseness=1
%
\begin{eqnarray}
\mathbf{z}_{i} &\sim&\mathrm{N}(\mathbf{z}_{i} \mid\mathbf{0},
\Sigma
_{\theta_i}), \\
\bolds{\theta_i} &\sim& G,\qquad   G \sim\operatorname{DP}(\alpha,
G_0), \\
G_0 &=& \mathrm{N}(\phi\mid d, D) \times\operatorname{IG}(v \mid a,
b)   ,
\end{eqnarray}
where the hyperparameters $(d,D)$ and $(a,b)$ must be chosen to reflect
the expected properties of the base measure $G_0$ (which is a product
of two independent distributions, a normal and an inverse-gamma), and
where $\alpha$ controls the degree of expected departure from the base measure.

Dirichlet-process priors for nonparametric Bayesian density estimation
were popularized by \citet{ferguson1973}, \citet{antoniak1974}, and
\citet{escobarwest1995}. An example of their use in analyzing
nonlinear autoregressive time series can be found in \citet{mullerDPM1997}.

Realizations of the Dirichlet process are almost surely discrete
distributions, and so we expect some of the $\bolds{\theta}_i$'s
to be the same across companies. This is the DP framework's primary
strength here, since it will facilitate borrowing of information across
time series. Simply allowing each time series to have its own $\phi$
and own $v$ would make for a simpler model (albeit with more
parameters), but the DP prior reflects the subject-specific knowledge
that significant clustering of autoregressive parameters should be expected.

This will lead to behavior similar to that predicted by a finite
mixture of AR(1) models, such as the kind considered by
Fr\"uhwirth-Schnatter and Kaufmann (\citeyear{fruhschatt2008}).
The Dirichlet-process prior, however, avoids the
complicated task of directly computing marginal likelihoods for mixture
models of different sizes, and so makes computation much simpler. Note
that since the marginal distribution of one draw from a
Dirichlet-process mixture depends only on the base measure, the DP acts
like a mixture model that is predictively matched to a single observation.

It is important to consider, of course, how choices for $\alpha$ and
$G_0$ affect the implied prior distributions both for the number of
mixture components and for the parameters associated with each
component. The marginal prior for the parameters of each mixture
component is simply given by the base measure, while the prior for
$\alpha$ can be described in terms of $n$ and $k$, the desired number
of mixture components, using results from \citet{antoniak1974}.


\section{A nonparametric alternative model}
\label{DPGP}

\subsection{Trajectories as random functions}

$\!$Section \ref{basicmodel} considered a simple constant-mean AR(1) model
for nonnull units, and Section \ref{DPAR} modified the AR(1)
assumption to account for a richer autoregressive structure. This
section now modifies the constant-mean assumption to allow for
time-varying nonzero trajectories upon which the autoregressive
residuals are superimposed. Most management teams, after all, do not
stay the same for 40 or 50 years, and we should not expect their
performance to stay the same, either.

Firm performance trajectories can be viewed as continuous random
functions that are observed at discrete (in this case, annual)
intervals. This is essentially a nonparametric version of a
mixed-effects model for longitudinal data [\citet
{kleinmanibrahim1998}]. Recent examples of such models include \citet
{bigelowdunson2005}, \citet{dunsonherring2007}, and, in a spatial
context, \citet{gelfandkottas2005}.

Let $\mathbf{f}_i = \{y_i(t), t \in\mathbb{R}^+\}$ be a
continuous-time stochastic process for each observed unit, and let
$\mathbf{t}_i$ denote the vector of times at which each unit was
observed. Then the model is
%
\begin{eqnarray}
\mathbf{y}_i &=& \mathbf{f}_i (\mathbf{t}_i) + \mathbf{z}_i, \label
{NPalt1} \\
\mathbf{z}_i &\sim&\mathrm{N}(\mathbf{z}_{i} \mid\mathbf{0},
\Sigma
_{\theta_i}), \qquad   \bolds{\theta_i} \sim G, \label{NPalt2} \\
\mathbf{f}_i &\sim& p \cdot F + (1-p) \cdot\delta_{F_0} \label
{NPalt3}   ,
\end{eqnarray}
where $G$ is the nonparametric residual model defined in Section \ref
{DPAR}, $\delta_{F_0}$ represents a point mass at the zero function
$F_0(t) = 0$, and $F$ is a random distribution over a function space
$\Omega$. As in Section \ref{basicmodel}, $p$ is the unknown prior
probability of coming from the alternative model $M_A$, represented in
this case by the distribution $F$. It is convenient to represent each
hypothesis test using a model index parameter $\gamma_i$: $\gamma_i =
0$ if $\mathbf{f}_i = F_0$ (i.e.,~the null model $M_0$ is true for
unit $i$), and $\gamma_i = 1$ otherwise.

\subsection{Choice of features for $M_A$}

The crucial consideration in using the above model for hypothesis
testing is that the space $\Omega$ from which each $\mathbf{f}_i$ is
drawn must be restricted to a sufficiently small class of functions.
This would be necessary even if $F$ were only being estimated, and not
tested against a simpler model: if $\Omega$ is too broad, then the
alternative model itself will not be likelihood-identified, since any
pattern of residuals could equally well be absorbed by the mean function.

But this guideline is even more important in model-selection problems;
an over-broad class of functions will mean that the random distribution
$F$ is vague, in the sense that the predictive distribution of
observables will be diffuse. It is widely known that using vague priors
for model selection can be produce very misleading results, and will
typically have the unintended consequence of sending the Bayes factor
in favor of the simpler model to infinity. This is often known as
Bartlett's paradox in the simple context of testing normal means [\citet
{bartlett1957}], but the same principle applies here.

It may also be the case that elements of $\Omega$ depend upon some
parameter $\bolds{\eta}$. Since this parameter appears only in
the alternative model, $\bolds{\eta}$ needs a proper prior, or
else the marginal likelihoods will be defined only up to an arbitrary
multiplicative constant.

Similar challenges occur in all model-selection problems. General
approaches and guidelines for choosing priors on nonshared parameters
can be found in \citet{laudibrahim1995}, \citet{ohagan1995}, \citet
{bergerpericchi1996}, and \citet{bergervarsh1998}. But very few tools
of analogous generality have been developed for nonparametric problems,
with most work concentrating on how to compute Bayes factors for
pre-specified models [\citet{basuchib2003}], or how to test a
parametric null against a nonparametric alternative of a suitably
restricted form [\citet{bergergug2001}].

This leaves just two obvious criteria for choosing $\Omega$ and $F$ in
the face of weak prior information:
\begin{enumerate}
\item Elements of $\Omega$ should be smooth, that is,~slowly varying
on the unit-time scale of the residual model. This will allow
deconvolution of the mean process from the residual, and reflects the
prior belief that the mean function will describe long-term departures
from $0$ in the face of short-term autoregressive jitters. (Indeed,
these departures are precisely what the methodology is meant to detect.)
\item$F$ should be centered at the null model, and should concentrate
most of its mass on elements of $\Omega$ that predict $\mathbf{y}$
values on a scale similar to those predicted by the null model. This
will avoid Bartlett's paradox, and generalizes the argument made by
\citet{jeffreys1961} in recommending an appropriately scaled Cauchy
prior for testing normal means.
\end{enumerate}

These criteria allow much wiggle room, but at least provide a starting
point. Unfortunately there is no objective solution, in this or in any
model-selection problem, though the closest thing to a default approach
is to simply choose the marginal variance of the alternative process to
exactly match the marginal variance of the null process. Best, of
course, is to conduct a robustness study, where the features of the
nonparametric alternative not shared by the null are varied in order to
assess changes in the conclusions. This will usually be quite difficult
in large multiple-testing problems, since computations for just a
single version of the alternative model may be expensive.

The choice of $\alpha$, the precision parameter for the residual
Dirichlet-process prior, is relatively free by comparison, since this
parameter appears in both the null and alternative models. Strictly
speaking, in order to use a\break noninformative prior for $\alpha$,
verification of the conditions in\break \citet{bergervarsh1998} regarding
group invariance is necessary, which is difficult in this case. (The
issue is that a parameter does not necessarily mean the same thing in
both $M_0$ and $M_A$ just because it is assigned the same symbol in
each.) In the absence of a formal justification for using a
noninformative prior, the conservative approach is to elicit priors for
$\alpha$ in terms of the expected number of AR(1) mixture components
in each of $M_0$ and~$M_A$. Often there will be extrinsic justification
for choosing $\alpha$ to be the same under both models.

\section{A model for the corporate-performance data}
\label{examplemodel}

\subsection{Model details}

As an example of how tests involving (\ref{NPalt1})--(\ref{NPalt3})
can be constructed, this section outlines a nonparametric model for a
larger subset of the corporate-performance data containing 5498 firms.
This contains every publicly traded American company between 1965 and
2005 for which at least 15 years of history were available.

The class of Gaussian processes with some known covariance function is
ideally suited for modeling nonzero trajectories, since the covariance
function can be chosen to yield smooth functions with probability 1,
and since the prior marginal variance of the process can be controlled
exactly (so that Bartlett's paradox may be easily avoided). Gaussian
processes have the added advantage of analytical tractability, which is
very important in hypothesis testing because of the need to evaluate
the marginal likelihood of the data under the alternative model. More
general classes of functions are certainly possible, though perhaps
computationally challenging in the face of massive multiplicity.

One additional feature to account for is clustering, since management
theorists are interested in identifying a small collection of
archetypal trajectories that may correspond to different sources of
competitive advantage. Partitioning of firms into shared trajectories
is especially relevant for advocates of the so-called ``resource-based
view'' of the firm [\citet{wernerfelt1984}]. Additionally, clustering
on treatment effects is known to increase power in multiple-testing
problems [\citet{dahlnewton2007}].

The approach considered here is similar to that introduced by\break \citet
{dunsonherring2007}. The distribution of nonzero random functions is
modeled with a functional Dirichlet process:
%
\begin{eqnarray}
F &\sim&\operatorname{FDP}(\nu, \operatorname{GP}(C_{\bolds{\kappa
}})), \label
{FDPGP} \\
C_{\bolds{\kappa}}(t_1, t_2) &=& \kappa_1 \cdot\exp\biggl(
-0.5 \cdot\frac{|t_1 - t_2|}{\kappa_2} \biggr)^2 \label{FDPGP2}   .
\end{eqnarray}
The functional Dirichlet process in (\ref{FDPGP}) has precision
parameter $\nu$ and is centered at a Gaussian process with covariance
function $C_{\bolds{\kappa}}$. At time $t$, the value of the
function $\mathbf{f}_i$ has a Dirichlet-process marginal distribution:
$\mathbf{f}_i(t) \sim F(t)$, where $F(t) \sim\operatorname{DP}(\nu,
\mathrm{N}(0, \kappa
_1))$. In choosing the hyperparameter $\bolds{\kappa}$, close
attention must be paid to the marginal variance of the residual model,
so that variance inflation in (\ref{FDPGP2}) does not overwhelm the
Bayes factor. For greater detail on Gaussian processes for
nonparametric function estimation, see \citet{rasmussen2006}.

This model is significantly richer than the simplistic framework
developed in Section \ref{basicmodel}, but is similar in two crucial ways:
\begin{itemize}
\item[\textit{Centering at the null model},] since the Gaussian process in
(\ref{FDPGP2}) leads to $\mathrm{E}(\mathbf{f}_i \mid \gamma_i = 1) =
\mathbf{0}$. As before, it is equally likely a priori that a
firm's trajectory will be predominantly negative or predominantly positive.
\item[\textit{Variance inflation}] under the alternative model is controlled
through the choice of a single hyperparameter, with $\kappa_1$ in
(\ref{FDPGP2}) playing the role of $\sigma^2$ in (\ref{simple3}).
Hence, despite the complicated nonparametric wrapper, the
Ockham's-razor effect upon the implied marginal likelihoods still
happens in the familiar way.
\end{itemize}
The model also solves both of the major problems encountered in the ROA
data: time-varying nonzero trajectories, and clustering both of
trajectories and of company-specific parameters for the autoregressive residual.

An extensive prior-elicitation process was undertaken with three
experts in management theory who had originally compiled the data. For
the base measure of the Dirichlet-process mixture of AR(1) covariance
models, the same hyperparameters from the parametric model in (\ref
{basicprior1}) and (\ref{basicprior2}) were used. Hence, the starting
point for this elicitation was $\kappa_1 \approx1.94$, which is the
prior marginal variance of the residual AR process (assessed by
simulation), and $\kappa_2 = 15$ (on a 41-year time scale), which
reflected the experts' judgments about the long-term effects of
strategic choices made by firms. The elicitees were repeatedly shown
trajectories drawn from this prior and other similar priors, and soon
settled upon $\kappa_1 = 1.25$ and $\kappa_2 = 13$ as values that
better reflected their expectations. Additionally, they chose $\alpha=
10/\log N$, and $\nu= 15 / \log N$ on the basis of how many clusters
they expected.

For the actual data, a third trajectory model was also introduced: a DP
mixture of constant trajectories rather than of Gaussian-process
trajectories. This entails only a slight complication of the analysis,
in that now (\ref{NPalt3}) is a three-component mixture rather than a
two-component mixture. This is equivalent to including the
limiting-linear-model framework of \citet{gramacy2005}---whereby a
flat trajectory is given nonzero probability as an explicit limiting
case of the Gaussian process---inside the base measure of the
functional Dirichlet process itself.

\subsection{Results}
\label{exampleresults}

The above model was implemented using the blocked Gibbs-sampling
algorithm of \citet{ishwaranjames2001} to draw from the nonparametric
distributions $F$ and $G$. Convergence was assessed through multiple
restarts from different starting points, and was judged to be
satisfactory. The software is available from the journal website in a
supplementary file [Scott (\citeyear{scott2009b})].

Overall, 981 of 5498 firms were flagged as being from the alternative
model with greater than $50\%$ probability, representing an overall
discovery rate of about $18\%$. Of these, only 196 firms were from the
alternative model with greater than $90\%$ probability.

To assess robustness to the hyperparameter choices $\kappa_1$ and
$\kappa_2$, which control the marginal variance and temporal range of
of the Gaussian process base measure in (\ref{FDPGP2}), the results
were recomputed for a coarse grid of 12 pairs of values spanning $0.75
\leq\kappa_1 \leq2.25$ and $5 \leq\kappa_2 \leq20$. This
reflected the lower and upper ends of what the collaborating management
theorists considered reasonable on the basis of observing draws from
these priors.

As expected, larger values of $\kappa_1$ tended to yield fewer
nonnull classifications (due to variance inflation in the marginal
likelihoods), while larger values of $\kappa_2$ tended to punish firms
whose peaks and valleys in performance were short-lived over the
41-year time horizon. Many firms that were borderline in the original
analysis (that is, having $p_i$ just barely larger than $50\%$) were
reclassified as ``noise'' for certain other values of $\boldsymbol
{\kappa}$. Yet a stable cohort of $246$ firms were flagged as nonnull
in all 12 analyses, suggesting a reasonable degree of robustness with
respect to hyperparameter choice.

\begin{figure}

\includegraphics{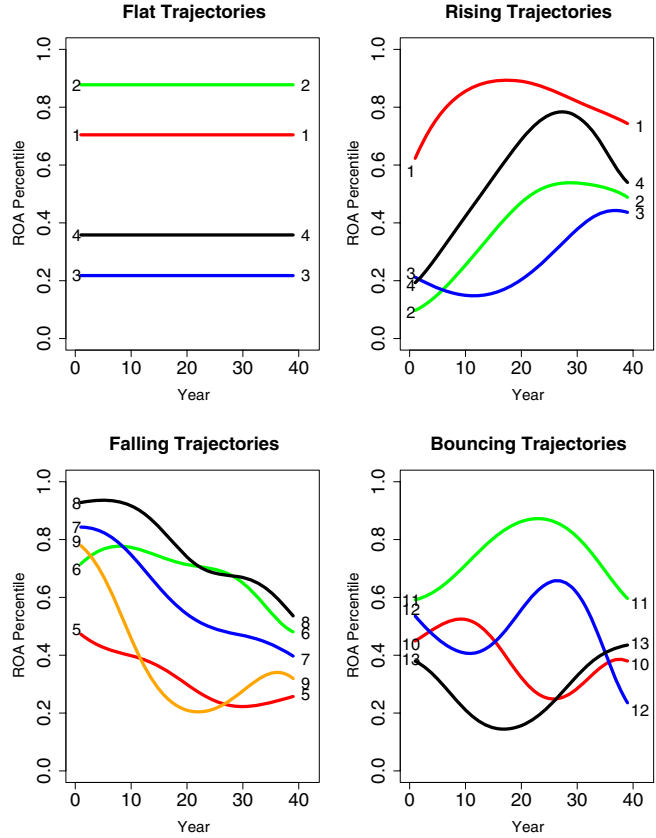}

\caption{The 17 highest-weight trajectories in the MLE estimate of the
alternative model, split into four loose categories. The $y$-axis is
given on the $\mathrm{N}(0,1)$ inverse-cdf scale to reflect the
quantile of
each firm's performance.}
\label{MLEtrajectories}
\end{figure}

The behavior of individual firms was then characterized by using the
MCMC history from the original analysis to get a maximum-likelihood
estimate of the nonparametric alternative model. The almost-sure
discreteness of the Dirichlet process means that this estimate is a
mixture of a small number of flat and Gaussian-process trajectories.
The 17 highest-weight trajectories in the MLE are in Figure \ref
{MLEtrajectories}; they are split into four loose categories reflecting
different archetypes of company performance.

\begin{table}
\caption{Posterior probabilities for six
firms that were flagged as being from trajectory GP 8 with greater than
$50\%$ probability in the MLE clustering analysis; see Figure \protect\ref
{MLEtrajectories} for labels}\label{fallersummaries}
\begin{tabular*}{\textwidth}{@{\extracolsep{\fill}}lld{2.0}d{2.0}d{2.0}ccc@{}}
\hline
\textbf{GV} & \textbf{Company name} & \multicolumn{1}{c}{\textbf{Flat 1}} & \multicolumn{1}{c}{\textbf{Flat 2}} & \multicolumn{1}{c}{\textbf{GP 7}} & \multicolumn{1}{c}{\textbf{GP 8}}
 & \multicolumn{1}{c}{\textbf{Other}} & \multicolumn{1}{c@{}}{\textbf{Null}} \\
\hline
11535 & Winn-Dixie Stores & 4 & 1 & 10 & 78 & 5 & 2 \\
\phantom{0}4828 & Delhaize America & 7 & 2 & 11 & 75 & 4 & 1 \\
\phantom{0}6830 & Lubrizol Corp & 20 & 9 & 2 & 64 & 4 & 1 \\
\phantom{0}7139 & Maytag Corp & 25 & 2 & 8 & 57 & 2 & 6 \\
\phantom{0}4323 & Emery Air Freight & 5 & 2 & 29 & 57 & 4 & 3 \\
\phantom{0}7734 & National Gas \& Oil & 18 & 13 & 11 & 53 & 3 & 2 \\
\hline
\end{tabular*}
 \tabnotetext[]{ta}{GV refers to a unique corporate identifier.}
\end{table}

\begin{figure}

\includegraphics{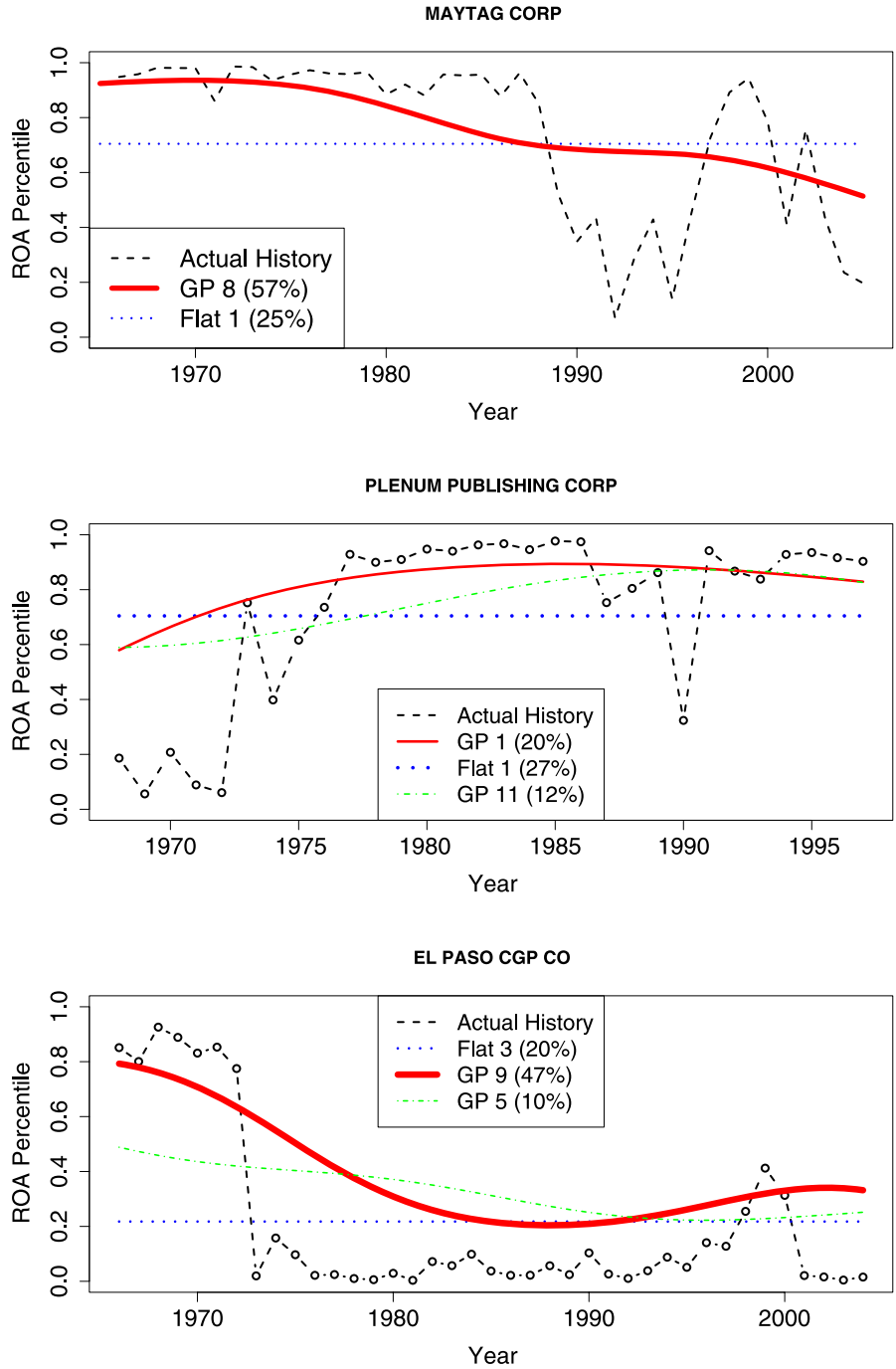}

\caption{Maytag, Plenum Publishing and El Paso CGP: actual ROA
histories along with trajectory-membership probabilities from the MLE
clustering analysis.}
\label{threeexamplefirms}
\end{figure}

This allows an MLE clustering analysis: if $F$ in (\ref{FDPGP}) is
frozen at the mixture model in Figure \ref{MLEtrajectories} and the
MCMC rerun, it is possible to flag companies that come from specific
clusters. (Strictly speaking, only the first 17 atoms in the
stick-breaking approximation of $F$ were frozen; others atoms were
still considered, but they were allowed to vary.) Examples of the kinds
of summaries available are in Table \ref{fallersummaries} and Figure
\ref{threeexamplefirms}.

Some general features of the methodology are apparent from these results:

\begin{itemize}
\item There is substantial shrinkage of estimated mean trajectories
back toward the global average (i.e.,~the 50th percentile). This is
itself a form of multiplicity correction, in that extreme outcomes are
quite likely to be attributed to chance even among those firms flagged
as being from the alternative model.
\item Often there is no dominant trajectory in the MLE cluster set that
characterizes a specific firm's history. For the three firms in Figure
\ref{threeexamplefirms}, the probability is split among two or even
three trajectories.
\item Model-averaged predictions of future performance are available
with very little extra work, since full MCMC histories of each firm's
trajectory and residual model are available.
\item The MLE clustering analysis can provide evidence for historical
evolution within specific firms, which is of great interest as the
subject of follow-up case studies. Maytag, for example, displays
markedly different performance patterns before and after 1987, which is
reflected in its high probability of being from a falling trajectory
(GP 8).
\end{itemize}

It must be emphasized that any such clustering analysis is at best an
approximation, aside from the fact that the models themselves are also
approximations. It relies, after all, upon a single point estimate of
the trajectories composing the random distribution $F$, and as such
ignores uncertainty about the trajectories themselves. In the example
at hand, several independent runs were conducted; each yielded a
different MLE cluster set, but the same broad patterns (e.g.,~something
like GP~8, something like Flat 3, and so on) emerged each time,
suggesting at least some degree of robustness of the qualitative conclusions.

Still, the most reliable quantities are the inclusion probabilities $\{
p_i\}$ computed from the full nonparametric analysis, which should form
the basis of claims regarding which units are from the null and which
are not.

\section{Type-I error performance: a simulation study}
\label{simstudy}

This section recapitulates the simulation study of Section \ref
{prelimrobustnessstudy} using the more complicated models. The goal is
to assess Type-I error performance by applying the methodology outlined
here to a simulated data set where the number of nonzero trajectories
is known.

The true residual model $G$ was constructed using a single MCMC draw
from the stick-breaking representation of the nonparametric residual
model for the corporate-performance data. This corresponded roughly to
a 21-component mixture of AR(1) models (since 39 of the 60 atoms in the
stick-breaking representation of $G$ had trivial weights), with many
$(\phi, v)$ pairs that differed starkly in character.

To simulate the true mean trajectories, two independent sets of $5500$
draws were taken from the prior in (\ref{NPalt1})--(\ref{NPalt3}). In
the first simulated set of trajectories, the true value of $p$ was
fixed at $1/5$ in order to roughly approximate the fraction of
discoveries on the ROA data. In the second set, $p$ was fixed at
$1/55$, to reflect a much sparser collection of signals. Upon sampling,
these yielded $1126$ and $98$ nonzero trajectories, respectively.

In both studies, each of these trajectories was convolved with a single
independent vector of autoregressive noise drawn from the true $G$---in
other words, with a highly complex pattern of residual variation of the
type that was shown to flummox the ``single AR(1)'' model of Section
\ref{basicmodel}. For the sake of comparison, the same set of $5500$
noise vectors was used for each experiment.

The results of these simulations were very encouraging for the Type-I
error performance of the more complicated model. In the first simulated
data set, $297$ of the $1126$ nonzero trajectories were flagged as
nonnulls with greater than $50\%$ probability. Only 24 of the 4374
null companies were falsely flagged, and of these, only $3$ had larger
than a $70\%$ inclusion probability.

In the second simulated data set, of the $98$ known nonzero
trajectories, $24$ had posterior inclusion probabilities greater than
$50\%$, and $6$ had inclusion probabilities greater than $90\%$. Of the
$5402$ null trajectories, only $2$ had inclusion probabilities greater
than $50\%$ (these two were only $63\%$ and $67\%$).

If $p_i \geq0.5$ is taken as the decision rule for a
posteriori classification of a trajectory as nonnull (which again
reflects a symmetric 0--1 loss function), then the realized
false-discovery rates were only $7.7\%$ on 26 discoveries in the
sparser case, and only $7.5\%$ on 321 discoveries in the denser case.
(The closeness of these two realized false-discovery rates may simply
be a coincidence of the particular values of $p$ chosen).

This suggests that a sizeable fraction of the $981$ firms flagged as
nonnull trajectories in Section \ref{exampleresults} represent
nonaverage performers, and are not false positives.

\section{Summary}
\label{summary}

This paper has described a framework for Bayesian multiple hypothesis
testing in time-series analysis. The proposed methodology requires
specifying only a few key hyperparameters for the nonparametric null
and alternative models, and general guidelines for choosing these
quantities have been given. Importantly, no ad-hoc ``correction
factors'' are necessary in order to introduce a penalty for multiple
testing. Rather, this penalty happens quite naturally by treating the
prior inclusion probability $p$ as an unknown quantity with a
noninformative prior.

Na\"ive characterizations of the null hypothesis are shown to have poor
error performance, suggesting that the Bayesian procedure is highly
sensitive to the accuracy of the null model used to describe an
``average'' time series. Yet once a sufficiently complex model for
residual variation is specified, the procedure exhibits very strong
control over the number of false-positive declarations, even in the
face of firms with different autoregressive profiles. The difference
between the results in Section \ref{prelimrobustnessstudy} and Section
\ref{simstudy} highlights the effectiveness and practicality of using
nonparametric methods as a general error-robustification tactic in
multiple-testing problems.

Posterior inference for a specific time series can be summarized in at
least three ways: by quoting $p_i$ (the probability of that unit's
being from the alternative model), by performing an MLE clustering
analysis as in Section \ref{exampleresults}, or by plotting the
posterior draws of the unit's nonzero mean trajectory. Plots such as
Figure \ref{threeexamplefirms} can be quite useful for communicating
inferences to nonexperts, as in the management-theory example
considered throughout.

The companies flagged as impressive performers, of course, can only be
judged so with respect to a particular notion of randomness: the DP
mixture of autoregressive models described in Section \ref{DPAR}.
Inference on the nonzero trajectories can still reflect model
misspecification, and cannot unambiguously identify companies that have
found a source of sustained competitive advantage. This
Dirichlet-process model, however, is a much more general statement of
the null models postulated by \citet{denrell2003} and \citet
{denrell2005}, suggesting that the procedure described here can
identify firms that, with high posterior probability, depart from
randomness in a specific way that may be interesting to researchers in
strategic management.

\begin{supplement}
\sname{Supplement}
\stitle{\phantom{0}DPARtestingAoAS.zip}
\slink[doi]{10.1214/09-AOAS252SUPP}
\slink[url]{http://lib.stat.cmu.edu/aoas/252/supplement.zip}
\sdatatype{.zip}
\sdescription{The data on corporate performance described in this
paper is freely available for those with access to Standard and Poor's
Compustat database. Annual return on assets is computed as (net
income)$/$(total assets), which are Compustat codes NI and AT,
respectively. ROA was further adjusted by regressing upon year, GICS
industry codes, debt-to-equity ratio and sales, all of which are also
available on Compustat. A simulated data set and the software necessary
to implement these models are freely available in the supplemental file
entitled ``DPARtestingAoAS.zip.''}
\end{supplement}

\section*{Acknowledgments}
 The author would like to thank Mumtaz Ahmed,
Michael Raynor, Lige Shao, Jim Guszcza and Jim Wappler of Deloitte
Consulting for their insight into the application discussed here, along
with Andy Henderson of the University of Texas and Carlos Carvalho of
the University of Chicago for all their helpful feedback.
%

\printaddresses


\begin{thebibliography}{99}
\bibitem[\protect\citeauthoryear{Antoniak}{1974}]{antoniak1974}
\textsc{Antoniak, C.} (1974). Mixtures of {D}irichlet processes with
applications to {B}ayesian nonparametric problems. \textit{Ann.
Statist.} \textbf{2} 1152--1174.
\MR{0365969}

\bibitem[\protect\citeauthoryear{Bartlett}{1957}]{bartlett1957}
\textsc{Bartlett, M.} (1957). A comment on {D}.{V}.~{L}indley's statistical
paradox. \textit{Biometrika} \textbf{44} 533--534.
\MR{0086727}

\bibitem[\protect\citeauthoryear{Basu and Chib}{2003}]{basuchib2003}
\textsc{Basu, S.} and \textsc{Chib, S.} (2003). Marginal likelihood
and {B}ayes factors
for {D}irichlet process mixture models. \textit{J. Amer. Statist.
Assoc.} \textbf{98} 224--235.
\MR{1965688}

\bibitem[\protect\citeauthoryear{Berger, Pericchi and
Varshavsky}{1998}]{bergervarsh1998}
\textsc{Berger, J., Pericchi, L.} and \textsc{Varshavsky, J.} (1998).
{B}ayes factors
and marginal distributions in invariant situations. \textit{Sankhya Ser.~A}
\textbf{60} 307--321.
\MR{1718789}

\bibitem[\protect\citeauthoryear{Berger and Guglielmi}{2001}]{bergergug2001}
\textsc{Berger, J.~O.} and \textsc{Guglielmi, A.} (2001). {B}ayesian
and conditional
frequentist testing of parametric model versus nonparametric alternatives.
\textit{J. Amer. Statist. Assoc.} \textbf{96} 174--184.
\MR{1952730}

\bibitem[\protect\citeauthoryear{Berger and
Pericchi}{1996}]{bergerpericchi1996}
\textsc{Berger, J.~O.} and \textsc{Pericchi, L.} (1996). The
intrinsic {B}ayes factor
for model selection and prediction. \textit{J. Amer. Statist. Assoc.}
\textbf{91} 109--122.
\MR{1394065}

\bibitem[\protect\citeauthoryear{Bigelow and
Dunson}{2005}]{bigelowdunson2005}
\textsc{Bigelow, J.} and \textsc{Dunson, D.} (2005). Semiparametric
classification in
hierarchical functional data analysis. Technical report, Duke Univ.,
Dept. Statistical Science.

\bibitem[\protect\citeauthoryear{Bowman and
Helfat}{2001}]{bowmanhelfat2001}
\textsc{Bowman, E.~H.} and \textsc{Helfat, C.~E.} (2001). Does
corporate strategy
matter? \textit{Strategic Management Journal} \textbf{22} 1--23.

\bibitem[\protect\citeauthoryear{Carvalho and
Scott}{2009}]{carvalhoscott2007}
\textsc{Carvalho, C.~M.} and \textsc{Scott, J.~G.} (2009). Objective
{B}ayesian model
selection in {G}aussian graphical models. \textit{Biometrika} \textbf{96} 497--512.

\bibitem[\protect\citeauthoryear{Cui and George}{2008}]{cuigeorge2006}
\textsc{Cui, W.} and \textsc{George, E.~I.} (2008). Empirical {B}ayes
vs.~fully {B}ayes
variable selection. \textit{J. Statist. Plann. Inference}
\textbf{138} 888--900.
\MR{2416869}

\bibitem[\protect\citeauthoryear{Dahl and Newton}{2007}]{dahlnewton2007}
\textsc{Dahl, D.~B.} and \textsc{Newton, M.~A.} (2007). Multiple
hypothesis testing by
clustering treatment effects. \textit{J. Amer. Statist. Assoc.}
\textbf{102} 517--526.
\MR{2325114}

\bibitem[\protect\citeauthoryear{Denrell}{2003}]{denrell2003}
\textsc{Denrell, J.} (2003). Vicarious learning, undersampling of
failure, and
the myths of management. \textit{Organizational Science} \textbf{4}
227--243.

\bibitem[\protect\citeauthoryear{Denrell}{2005}]{denrell2005}
\textsc{Denrell, J.} (2005). Selection bias and the perils of benchmarking.
\textit{Harvard Business Review} \textbf{April, 2005} 114--119.

\bibitem[\protect\citeauthoryear{Do, Muller and Tang}{2005}]{domuller2005}
\textsc{Do, K.-A., Muller, P.} and \textsc{Tang, F.} (2005). A
{B}ayesian mixture
model for differential gene expression. \textit{J. Roy.
Statist.
Soc. Ser. C} \textbf{54} 627--644.
\MR{2137258}

\bibitem[\protect\citeauthoryear{Dunson and
Herring}{2006}]{dunsonherring2007}
\textsc{Dunson, D.} and \textsc{Herring, A.} (2006). Semiparametric
{B}ayesian latent
trajectory models. Technical report, Duke Univ., Dept. Statistical
Science.

\bibitem[\protect\citeauthoryear{Escobar and West}{1995}]{escobarwest1995}
\textsc{Escobar, M.} and \textsc{West, M.} (1995). {B}ayesian density
estimation and
inference using mixtures. \textit{J.~Amer. Statist. Assoc.} \textbf
{90} 577--588.
\MR{1340510}

\bibitem[\protect\citeauthoryear{Ferguson}{1973}]{ferguson1973}
\textsc{Ferguson, T.} (1973). A {B}ayesian analysis of some nonparametric
problems. \textit{Ann. Statist.} \textbf{1} 209--230.
\MR{0350949}

\bibitem[\protect\citeauthoryear{Fr\"uhwirth-Schnatter and
Kaufmann}{2008}]{fruhschatt2008}
\textsc{Fr\"uhwirth-Schnatter, S.} and \textsc{Kaufmann, S.} (2008).
Model-based
clustering of multiple time series. \textit{J. Bus. Econom.
Statist.} \textbf{26} 78--89.
\MR{2422063}

\bibitem[\protect\citeauthoryear{Gelfand, Kottas and
MacEachern}{2005}]{gelfandkottas2005}
\textsc{Gelfand, A., Kottas, A.} and \textsc{MacEachern, S.} (2005).
{B}ayesian
nonparametric spatial modeling with Dirichlet process mixing. \textit
{J. Amer. Statist. Assoc.} \textbf{100} 1021--1035.
\MR{2201028}

\bibitem[\protect\citeauthoryear{George and Foster}{2000}]{georgefoster2000}
\textsc{George, E.~I.} and \textsc{Foster, D.~P.} (2000). Calibration
and empirical
{B}ayes variable selection. \textit{Biometrika} \textbf{87} 731--747.
\MR{1813972}

\bibitem[\protect\citeauthoryear{Gramacy}{2005}]{gramacy2005}
\textsc{Gramacy, R.} (2005). {B}ayesian treed {G}aussian process
models. Ph.D.
thesis, Univ. California--Santa Cruz.

\bibitem[\protect\citeauthoryear{Harrigan}{1985}]{harrigan1985}
\textsc{Harrigan, K.} (1985). An application of clustering for
strategic group
analysis. \textit{Strategic Management Journal} \textbf{6} 55--73.

\bibitem[\protect\citeauthoryear{Hawawini, Subramanian and
Verdin}{2003}]{hawawinietal2003}
\textsc{Hawawini, G., Subramanian, V.} and \textsc{Verdin, P.}
(2003). Is performance
driven by industry- or firm-specific factors? A new look at the evidence.
\textit{Strategic Management Journal} \textbf{24} 1--16.

\bibitem[\protect\citeauthoryear{Ishwaran and
James}{2001}]{ishwaranjames2001}
\textsc{Ishwaran, H.} and \textsc{James, L.} (2001). Gibbs sampling
methods for
stick-breaking priors. \textit{J. Amer. Statist. Assoc.} \textbf{96} 161--173.
\MR{1952729}

\bibitem[\protect\citeauthoryear{Jefferys and
Berger}{1992}]{jefferysberger92}
\textsc{Jefferys, W.} and \textsc{Berger, J.} (1992). Ockham's razor
and {B}ayesian
analysis. \textit{Amer. Sci.} \textbf{80} 64--72.

\bibitem[\protect\citeauthoryear{Jeffreys}{1961}]{jeffreys1961}
\textsc{Jeffreys, H.} (1961). \textit{Theory of Probability}, 3rd ed. Oxford
Univ.
Press.
\MR{0187257}

\bibitem[\protect\citeauthoryear{Johnstone and
Silverman}{2004}]{johnstonesilverman2004}
\textsc{Johnstone, I.~M.} and \textsc{Silverman, B.~W.} (2004).
Needles and Straw in
Haystacks: Empirical-{B}ayes estimates of possibly sparse sequences.
\textit{Ann. Statist.} \textbf{32} 1594--1649.
\MR{2089135}

\bibitem[\protect\citeauthoryear{Kleinman and
Ibrahim}{1998}]{kleinmanibrahim1998}
\textsc{Kleinman, K.} and \textsc{Ibrahim, J.} (1998). A
semiparametric {B}ayesian
approach to the random effects model. \textit{Biometrics} \textbf{54}
921--938.

\bibitem[\protect\citeauthoryear{Laud and Ibrahim}{1995}]{laudibrahim1995}
\textsc{Laud, P.} and \textsc{Ibrahim, J.} (1995). Predictive model selection.
\textit{J. Roy. Statist. Soc. Ser. B} \textbf{57} 247--262.
\MR{1325389}

\bibitem[\protect\citeauthoryear{M\"uller, West and
MacEachern}{1997}]{mullerDPM1997}
\textsc{M\"uller, P., West, M.} and \textsc{MacEachern, S.} (1997).
{B}ayesian models
for non-linear auto-regressions. \textit{J. Time Ser.
Anal.} \textbf{18}
593--614.
\MR{1603392}

\bibitem[\protect\citeauthoryear{O'Hagan}{1995}]{ohagan1995}
\textsc{O'Hagan, A.} (1995). Fractional {B}ayes factors for model comparison.
\textit{J. Roy. Statist. Soc. Ser. B} \textbf{57} 99--138.
\MR{1325379}

\bibitem[\protect\citeauthoryear{Rasmussen and
Williams}{2006}]{rasmussen2006}
\textsc{Rasmussen, C.~E.} and \textsc{Williams, C.} (2006). \textit
{Gaussian Processes for Machine
Learning}. MIT Press, Cambridge, MA.
\MR{2514435}

\bibitem[\protect\citeauthoryear{Ruefli and
Wiggins}{2000}]{ruefliwiggins2000}
\textsc{Ruefli, T.~W.} and \textsc{Wiggins, R.~R.} (2000).
Longitudinal performance
stratification: An iterative {K}olmogorov--{S}mirnov approach.
\textit{Management Sci.} \textbf{46} 685--692.

\bibitem[\protect\citeauthoryear{Ruefli and
Wiggins}{2002}]{ruefliwiggins2002}
\textsc{Ruefli, T.~W.} and \textsc{Wiggins, R.~R.} (2002). Sustained
competitive
advantage: Temporal dynamics and the incidence and persistence of
superior economic performance. \textit{Organization Science} \textbf
{13} 81--105.

\bibitem[\protect\citeauthoryear{Scott}{2009}]{scott2009b}
\textsc{Scott, J.~G.}~(2009).
Supplement to ``Nonparametric Bayesian multiple testing for
longitudinal performance stratification.''

\bibitem[\protect\citeauthoryear{Scott and Berger}{2006}]{scottberger06}
\textsc{Scott, J.~G.} and \textsc{Berger, J.~O.} (2006). An
exploration of aspects of
{B}ayesian multiple testing. \textit{J.~Statist. Plann.
Inference} \textbf{136} 2144--2162.
\MR{2235051}

\bibitem[\protect\citeauthoryear{Scott and Berger}{2008}]{scottberger2007}
\textsc{Scott, J.~G.} and \textsc{Berger, J.~O.} (2008). {B}ayes and
Empirical-{B}ayes
multiplicity adjustment in the variable-selection problem. Discussion Paper
2008-10, Duke Univ., Dept. Statistical Science.

\bibitem[\protect\citeauthoryear{Scott and
Carvalho}{2008}]{scottcarvalho2007b}
\textsc{Scott, J.~G.} and \textsc{Carvalho, C.~M.} (2008).
Feature-inclusion stochastic
search for {G}aussian graphical models. \textit{J.
Comput.
Graph. Statist.} \textbf{17} 790--808.

\bibitem[\protect\citeauthoryear{Wernerfelt}{1984}]{wernerfelt1984}
\textsc{Wernerfelt, B.} (1984). The resource-based view of the firm.
\textit{Strategic Management Journal} \textbf{5} 171--180.

\end{thebibliography}
\end{document}